\documentclass[fleqn,twoside]{article}
\usepackage{espcrc2}

\usepackage{graphicx}
\usepackage{epsf}
\usepackage[figuresright]{rotating}

    \newcommand{\tr}{\mbox{Tr}}
    \newcommand{\half}{\frac{1}{2}}

\newcommand{\AmS}{{\protect\the\textfont2
  A\kern-.1667em\lower.5ex\hbox{M}\kern-.125emS}}

\title{\vspace{-3.65cm}
       {\normalsize DESY 01-161}    \\[-0.2cm]
       {\normalsize HU-EP-01/42}   \\[-0.2cm]
       {\normalsize October 2001}   \\
       \vspace{2.72cm}
Continuum limit in abelian projected SU(2)
 lattice gauge theory \thanks{Poster presented by V. Bornyakov at 
{\it Latice 2001}, Berlin.}}

\author{V. Bornyakov
\address{
NIC/DESY Zeuthen, Platanenallee 6, 15738 Zeuthen, Germany  \\} 
\thanks{On leave of absence from IHEP, Protvino, Russia}
and M. M\"uller-Preussker
\address{
Humboldt-Universit\"at zu Berlin, Institut f\"ur Physik, 10115 Berlin, Germany}
}

\begin{document}

\begin{abstract}
We study the continuum limit of the abelian string tension and the 
density of abelian monopoles calculated after carefully fixing 
the maximal abelian gauge by employing the simulated annealing 
algorithm. We present the evidence that the abelian string tension 
converges to the nonabelian one in the continuum limit. For the 
monopole density we confirm earlier findings that the density of 
the properly defined infrared monopoles has correct scaling while 
the total density seems divergent in the continuum limit due to 
ultraviolate contributions. We also compare with results obtained 
with the usual iterative gauge fixing algorithm.
\vspace{.2pc}
\end{abstract}

\maketitle
\section{INTRODUCTION}
Numerous lattice studies have provided support for the dual 
superconductor scenario of confinement \cite{thm}. Most of the
results have been obtained in the maximal abelian gauge (MAG) at 
particular values of the lattice spacing. 
To our knowledge no attempt was made for extrapolation into 
the continuum limit. In this paper we want to fill in this gap. 

\section{GAUGE FIXING AND SIMULATION DETAILS}
We work with $SU(2)$ gluodynamics in the MAG which is defined by 
maximizing the functional \cite{klsw}
\begin{equation}
F(U^g) = \frac{1}{8L^4}\sum_{x,\mu}
\mbox{Tr} \left [ U^g_{x,\mu}\sigma_3 U_{x,\mu}^{g,\dagger}\sigma_3\right] 
\label{func}
\end{equation}
with respect to the gauge transformations $g_{x}:
U^g_{x,\mu}=g_{x}U_{x,\mu} g_{x+\mu}^\dagger$. The abelian projection 
$U_{x,\mu} \rightarrow u_{x,\mu}$ is defined by the following relations
\begin{equation}
U_{x,\mu}^{ii} = cos(\phi_{x,\mu}) e^{i\theta_{x,\mu}^i},~~~ 
\theta_{x,\mu}^1=-\theta_{x,\mu}^2 \equiv \theta_{x,\mu} 
\end{equation}
\begin{equation}
u_{x,\mu} \equiv $diag $\{ e^{i\theta_{x,\mu}},e^{-i\theta_{x,\mu}} \}.
\end{equation}
\begin{table}[hpt]\caption{}
\begin{tabular}{|c|c|c|c|} \hline
  & $\beta=2.4$ & $\beta=2.5$ & $\beta=2.6$ \\ \hline
$L=L_{s,t}$ & 24  &           24        & 28         \\ 
$a$ & 0.12 fm &   0.085 fm      & 0.06 fm    \\
$L \cdot a$& 2.9 fm & 2.1 fm     & 1.7 fm      \\ 
$<F>_{SA}$ & .7336(1)  & .7511(1)  & .7662(1)  \\
$<F>_{RO}$& .7310(1)  & .7491(1)  & .7648(1)  \\  \hline
\end{tabular}
\vspace{-.5cm}
\end{table}
It is well known that the functional (\ref{func}) has many local
maxima corresponding to Gribov copies \cite{grib}. The standard procedure  
consists in selecting a local maximum one finds more or less randomly by an
iterative algorithm. Our prescription of
gauge fixing is to find the global maximum  of (\ref{func}). 
This prescription resembles the minimal Landau gauge \cite{zwanziger}.
The 'global maxima' definition of gauge fixing is known to be
a limit of the global gauge fixing introduced in \cite{parjl}.
As a numerical tool we use the simulated annealing 
(SA) algorithm \cite{kgv,bbms}. This algorithm helps to get closer 
to the global maxima but at some low auxiliary temperature $T$ it, 
nevertheless, gets trapped into metastable states of the corresponding 
spin glass.  To overcome this problem we applied the procedure to 
every MC configuration 10 times by starting from different gauge 
copies and used the copy with maximal value of $F(U)$. For comparison we carried out gauge fixing with the standard
iterative algorithm with
alternating relaxation and over-relaxation sweeps (RO algorithm).
Only one gauge copy was considered as it is commonly done. 

The simulation details are presented in Table~1. There we
also show the average value of $F(U)$ for the two
algorithms. We investigated 20 statistically independent configurations
per $\beta$-value.

\section{MONOPOLE DENSITY }
We use the DeGrand-Toussaint definition of magnetic currents \cite{dt} 
\begin{equation}
 k_{x,\mu} = \frac {1}{4\pi} \varepsilon_{\mu \nu \rho
\sigma} \partial_{\nu} \overline{\theta}_{x,\rho \sigma}\quad 
\end{equation}
with
$~~ \overline{\theta}_{x,\nu \mu} 
    = \partial_{\mu} \theta_{x,\nu} 
    - \partial_{\nu} \theta_{x,\mu} + 2 \pi m_{x,\mu\nu}~~$ and \\ 
$-\pi \leq \overline{\theta}_{n,\mu\nu} < \pi, \quad
 m_{n,\mu\nu} = 0, \pm 1, \pm 2$\,.\\
The physical monopole density is then defined as
\begin{equation}
  \rho_{\mbox{\scriptsize mon}}=\frac{1}{4(aL_s)^3L_t}
  \langle\sum_{x,\mu}k_{x,\mu}\rangle \,. 
\end{equation}
Following \cite{bcpsvz} we will call monopoles relevant for confinement
`infrared' (IR) monopoles and the rest of monopoles, normally forming 
small loops, `ultraviolate' (UV) ones. The  
corresponding densities are denoted as $\rho_{\mbox{\scriptsize mon}}^{IR,UV}$.
In \cite{ht} it has been found that monopoles from the 
largest cluster (LC) are sufficient to reproduce the string
tension. This observation suggests that the IR monopoles can be identified
as the monopoles belonging to the LC \cite{bcpsvz}.
However, as it has been indicated  in \cite{ht} the LC  
may split into parts if the volume is not large enough. We
confirm this observation. Moreover, we have found that, when the LC splits,
then there usually appear clusters with nontrivial winding 
\begin{equation}
 w_\mu = \frac{1}{L_\mu}\sum_{k_{x,\mu}\in cluster} k_{x,\mu} \,.
\end{equation}
Therefore, we suggest the following definition of IR monopoles:
monopoles from LC plus those from clusters with nontrivial 
$w_\mu$.  
\begin{figure}[htb]
\hbox{
\epsfysize=5.0cm
\epsfxsize=7.5cm
\epsfbox{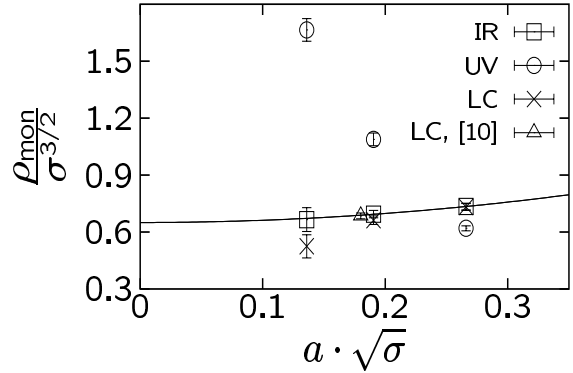}}
\vspace{-.8cm}
\caption{\label{dens}\it{Density of monopoles vs. lattice spacing.
\vspace{-.8cm}
}}
\end{figure}
In Fig. \ref{dens} we present our results for 
$\rho_{\mbox{\scriptsize mon}}^{IR,UV}$, measured according 
to the new definition. 
\footnote{
Throughout this work we use the following values of the nonabelian
string tension $\sqrt{\sigma a^2}$ \cite{fhk}: 0.266(2) at $\beta=2.4$;
0.1905(8) at $\beta=2.5$; 0.136(4) at $\beta=2.6$. We set the scale 
putting $\sqrt{\sigma}=440$ Mev.}
We also show our results for the density of monopoles belonging to the LC,
$\rho_{\mbox{\scriptsize mon}}^{LC}$, together with data from \cite{ht}.
We found that when the lattice size is large enough the clusters with
nontrivial windings are rare and the two definitions give coinciding 
results. This is also supported by the agreement between our 
$\rho_{\mbox{\scriptsize mon}}^{IR}$ value obtained at $\beta=2.5, 
L \cdot a=2.1$fm and the 
$\rho_{\mbox{\scriptsize mon}}^{LC}$ value from \cite{ht}
obtained at $\beta=2.5115, L \cdot a=2.8$ fm. 
We conclude that the new definition 
of IR monopoles can be successfully used on lattices with size
$L\cdot a \geq 1.7$ fm. It is to be checked, how  this definition
works on smaller lattices.
We extrapolated $\rho_{\mbox{\scriptsize mon}}^{IR}$
to the continuum limit using a quadratic fit (solid curve in Fig.~1). 
We find that $\rho_{\scriptsize mon}^{IR} = 0.65(2) \sigma^{3/2}$. 
Data in Fig.~1 show that while $\rho_{\scriptsize mon}^{IR}$
is finite, $\rho_{\scriptsize mon}^{UV}$ seems to diverge providing
a divergent total density.
Our data indicate that 
$\rho_{\mbox{\scriptsize mon}}^{UV}/\sigma^{3/2} \sim 1/\sqrt{\sigma a^2} $. 

\section{ABELIAN STATIC POTENTIAL AND STRING TENSION}
We introduce the abelian Wilson loop 
\begin{equation}
W_{ab}(C) = \half \tr\left(\prod_{l \in C} u_{l}\right)
\end{equation}
and the abelian static potential
\begin{equation}
V^T_{ab}(R)=\log\left(\frac{\langle W_{ab}(R,T)\rangle}
{\langle W_{ab}(R,T+1)\rangle}\right) .
\end{equation}
On-axes as well as $(0,1,1)$ off-axes directions were used. 
We apply smearing for spatial abelian links to 
improve the signal-to-noise ratio.  
$V^T_{ab}(R)$ has been fitted to a constant within the range 
$ 5\leq Ta \leq 8 $ to obtain $V_{ab}(R)$.
We employ a dimensionless form of the usual parametrization for the 
static potential 
\begin{equation}
\frac{V_{ab}(R)}{\sqrt{\sigma}}=\frac{V_0}{\sqrt{\sigma}}-
\frac{e}{R\sqrt{\sigma}}+\frac{\sigma_{ab}}{\sigma} R\sqrt{\sigma}\,. 
\end{equation}
Data points were fitted to this form in the range 
$0.5 < R \cdot \sqrt{\sigma} < 2.5$.
\begin{figure}[t]
\hbox{
\epsfysize=5.0cm
\epsfxsize=7.5cm
\epsfbox{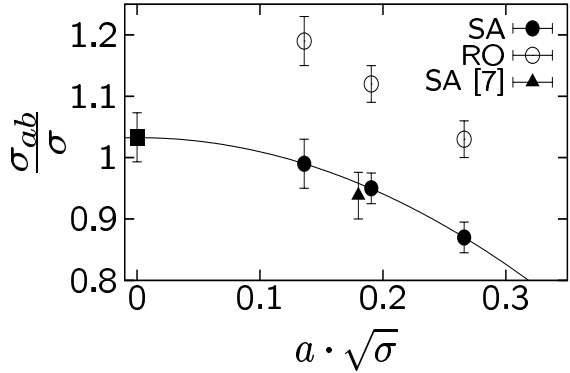}}
\vspace{-.8cm}
\caption{\label{sigma}\it Ratio of the abelian and nonabelian 
string tensions vs. 
lattice spacing for SA and RO algorithms. The solid line represents 
a quadratic fit. For comparison data from \cite{bbms} are shown.
}
\end{figure}
Fig. \ref{sigma} presents our main result. 
The continuum limit of $\sigma_{ab}$
agrees with that of $\sigma$ within error bars.
One can see a good agreement with a result from \cite{bbms} obtained
by the same method. It is worth noting
that our result is still biased by Gribov copies effects, although
these effects were substantially reduced in comparison with the common
procedure. Using results of \cite{bbms} we can estimate the bias
as $\Delta \sigma_{ab}/\sigma \sim -0.03$. Taking into account this
bias would not spoil our conclusion. 

We also made an attempt to compute the charge-two abelian static 
potential. In \cite{bbms,poul} it has been
conjectured that the charge-two abelian string tension 
$\sigma_{ab}^{q=2}$
is equal to the adjoint string tension. If this is true, then in the
continuum limit $\sigma_{ab}^{q=2}/\sigma \rightarrow 8/3$.
Unfortunately, our statistics appeared to be insufficient to provide a 
reliable estimation of $\sigma_{ab}^{q=2}$. 
The work on this and also on the monopole 
static potential is in progress.

\section{CONCLUSIONS}

\begin{itemize}
\item[--] With Gribov copies effects taken under better control we 
found that in 
the continuum limit $\sigma_{ab}/\sigma = 1.03(4)$
supporting abelian dominance for the string tension within the maximal
abelian gauge.
\item[--] The density of IR monopoles turns out finite in the  
continuum limit:  $\rho_{\scriptsize mon}^{IR} = 0.65(2) 
\sigma^{3/2}$. 
Thus, the average distance between monopoles can be estimated as 
$d_{mon}=1.15(1)/\sqrt{\sigma}=0.52(1)$ fm.
\end{itemize}

\noindent
This work is partially supported by INTAS grant 00-00111. V.B.
acknowledges support from RFBR grants 99-01230a and 01-02-17456.


\begin{thebibliography}{99}
\newcommand{\prd}[1]{Phys.~Rev.~{\bf D#1}\ }
\newcommand{\npb}[1]{Nucl.~Phys.~{\bf B#1}\ }
\newcommand{\plb}[1]{Phys.~Lett.~{\bf #1B}\ }
\newcommand{\pr}[1]{Phys.~Rep.~{\bf #1}\ }


\bibitem{thm} G.~'t~Hooft, in `High Energy Physics',
               Proceedings of the EPS Int. Conf.,
               Palermo 1975, ed. A.~Zichichi; 
               S.~Mandelstam, \pr{C23} (1976) 245.
\bibitem{klsw} A.S.~Kronfeld, M.L.~Laursen, G.~Schierholz and U.-J.~Wiese, 
	       Phys. Lett. {\bf 198B} (1987) 516.  
\bibitem{grib} V.N.~Gribov, \npb{139} (1978) 1.
\bibitem{zwanziger} D.~Zwanziger, \npb{412} (1994) 657.
\bibitem{parjl} C.~Parrinello and G.~Jona-Lasinio, Phys.\ Lett.\
               {\bf B251} (1990) 175; D.~Zwanziger, \npb{345} (1990) 461.
\bibitem{kgv}  S.~Kirkpatrick, C.D.~Gelatt Jr. and M.P.~Vecchi,
               Science {\bf 220} (1983) 671; V.~Cerny, 
               J.\ Opt.\ The.\ Appl.\ {\bf 45} (1985) 41.
\bibitem{bbms} G.S.~Bali, V.~Bornyakov, M.~M\"uller-Preussker 
               and  K.\ Schilling, \prd{54}(1996) 2863. 
\bibitem{dt}   T.A.~DeGrand and D.~Toussaint, Phys. Rev. {\bf D22} (1980) 2478.
\bibitem{bcpsvz} V. Bornyakov, M. Chernodub, F. Gubarev, M. Polikarpov, 
	       T. Suzuki, A. Veselov and V. Zakharov, hep-lat/0103032. 
\bibitem{ht}   A.~Hart and M.~Teper, 
	       \prd{58}(1998) 014504; \prd{60}(1999) 114506.
\bibitem{fhk} J.~Fingberg, U.M.~Heller and F.~Karsch, \npb{392}(1993) 493. 
\bibitem{poul} G.~Poulis, \prd{54} (1996) 6974.
\end{thebibliography}
\end{document}